Information Mechanics: Evidence, Hypothesis & Theory September 2022

John Laurence Haller Jr.

Abstract

A unifying theory is put forward that entropy is equal to action. The crowning derivation is based on information theoretic methods and uses our hypothesis that "particles move via the discrete Bernoulli Process." While this hypothesis matches special relativity for the mean value of a particle's location, the variance appears to open the possibility of new physics. Specifically, the variance is dependent on the square of the particle's velocity and since no reference frame is given, we ask if this velocity is absolute and measurable. To answer this question, we conducted an experiment to measure the jitter on a clock and found a combined 26 sigma three-spike signal in the Fourier Transform of the jitter's magnitude. These spikes are predicted by the hypothesis since the laboratory's velocity fluctuates due to the Earth revolving and rotating as the solar system moves through the Universe. When analyzing the phase and magnitude, the data suggests the velocity of the laboratory is a validation of the Planck collaboration which measured the same velocity of Earth in the Cosmic Microwave Background reference frame. Lastly, a second experiment is underway to ratify these findings.

## 1. Summary

The purpose of this report is neither a historical accounting, nor mathematical physics for that matter. Rather it is an intriguing, albeit incomplete, exploration of what is possible. For a compilation of efforts over the past century, Deutsch's take on the 2nd quantum revolution is inspiring [0].

Notwithstanding, we begin by looking back for context. Let us be reminded that almost 4 centuries ago, Galileo introduced the concept of relativity and set the framework for motion [1]. A century after that, Newton came forward with his Laws of Motion and stated that an inertial frame is one where his 1st Law was valid and proposed an absolute reference frame determined by the fixed stars [2]. Yet just over a century ago, Michaelson & Morley's experiment to measure the speed of light [3] and Einstein's theory of special relativity [4] helped put a nail in the coffin of Newton's absolute reference frame.

While these advances led to a period of enlightenment, challenges still exist in unifying the resulting general theory of relativity with quantum mechanics. Specifically, general relativity is continuous while quantum mechanics is quantized. While it is useful to study the many attempts to bridge quantum mechanics with relativity [5,6,7,8], these theories have yet to coalesce into a single agreement or provide experimental evidence of a testable hypotheses.

I propose that new avenues of research could be waiting for us if we are willing to consider a new perspective. While consistent with current experiments, an artifact from the hypothesis that "particles move via the Bernoulli Process" is only visible by asking different questions. For example, Michaelson & Morley measured the mean of the speed of light without any consideration of variance, nor did they conduct it for a full year.

When we look at the variance of space-time coordinates, we see the Bernoulli Process adds a term that is proportional to the square of the velocity. Since, no reference frame has been defined other than that of the laboratory, we ask, "What is that velocity relative to?"

This paradigm, to consider variance alongside the mean value, is commonplace today in many other disciplines. For example, a stock trader knows that risk (or variance) is just as important as return (or mean). Or closer to home, in information theory, the entropy of a Gaussian distribution is dependent on its variance, not its mean.

While Reif [9] and Chandrasekar [10] have eloquently brought the Bernoulli process into physics, we wanted to follow the scientific method and test it. Here, we present a novel experiment to investigate this hypothesis and, in doing, uncover things that others have not.

In this new experiment, we are able to accurately measure the velocity of the Cosmic Microwave Background reference frame, consistent with the measurement from the Planck Collaboration [11]. However, by doing so, we open up enhancements to our understanding of motion in that the clock has a predictable jitter term that varies as the Earth rotates, revolves and translates across the Universe, which means our time reference has its own reference.

What is additionally exciting about the hypothesis that particle's move via the Bernoulli Process is that using the equations outlined below, it is possible to derive that the self-information of a particle, or its "entropy", is equal to the time integral of the mass energy minus the Lagrangian, or the "action".

The major efforts underway to unify gravity and quantum mechanics might benefit from these insights, since entropy and action are each a cornerstone of different major disciplines within physics and their unification could yield unexpected rewards. All that is needed is to suspend our skepticism for the moment and consider a different perspective, and in turn carefully construct inquiries to further understand relativity.

For example, a few illustrations where entropy is equal to action, pulled from prior work [12,13] are:

- entropy is quantized to natural units, $\log(e)$
- an explanation for the 2 slit experiment and a view into the wave particle duality
- a split Gaussian wave packet contains $\log(e)$ of entropy
- a spin $s$ particle is $(2s+1)\log(e)$ of entropy
- the entropy and action of a black hole
- a special black hole with the reduced Planck mass dubbed a "dark particle" where its solutions to quantum diffusion are equal to its solutions to Friedmann's equation

While these ideas have taken decades to turn into cohesive story with evidence, it is time to enlist out the great community to debate, affirm, refute, replicate and expound on these ideas.

### 1.1. Setup

Let us review the experimental data. Then, a hypothesis which predicts the experimental sigma. Next, the velocity of the CMB reference frame is validated as we discuss implications that a time standard has a preferred frame where the jitter is minimized. Following up with a few derivations of the governing thesis that entropy is unified with action, we argue on both experimental and theoretical grounds that resources should be brought to bear on these findings. A second experiment which is predicted to show a similar outcome is also shared. Additional details are given in the methods section to ensure replicability by others.

## 2. Experiment

### 2.1. Experiment on a time standard – clock jitter

While more details can be found in the method section below, the gist is as follows: take two atomic clocks and measure the timestamps of zero crossings of each clock; take the derivative of each of the time stamps to obtain the periods, subtract the two channels (to double to signal), take another derivative of each (to remove drift), take the absolute value, and lastly compute the Fourier Transform (FT) for a select set of 161 frequencies, Figure 1.

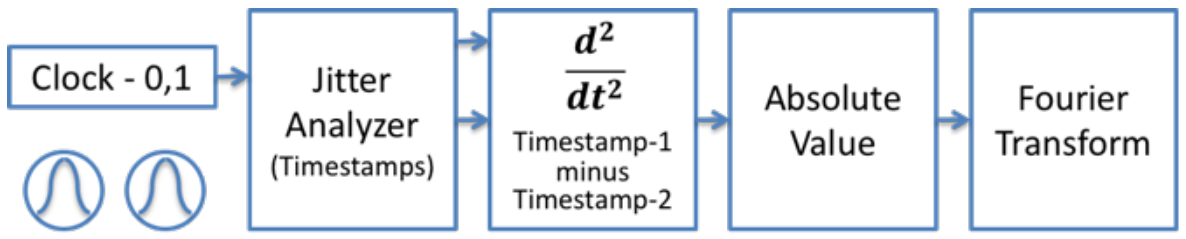

Figure 1 – A diagram of the experimental analysis

#### 2.1.1. Experimental Results

The magnitude of the resulting FT around one cycle per day with a precision of $\sim 1/(2.5 years)$ and an uptime (duty cycle) above 97% is shown in Figure 2. We see three spikes, each being 10+ standard deviations above the mean, with a combined 26 standard deviations. Since current theory does not predict any variation of the jitter with these periodicities, $\sigma_{theory} = 0$, the peaks of this graph are a measure of sigma.

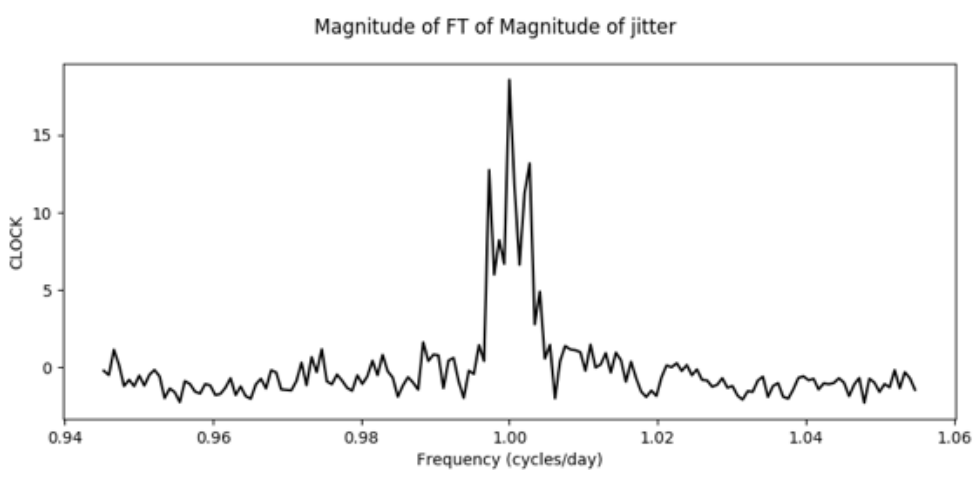

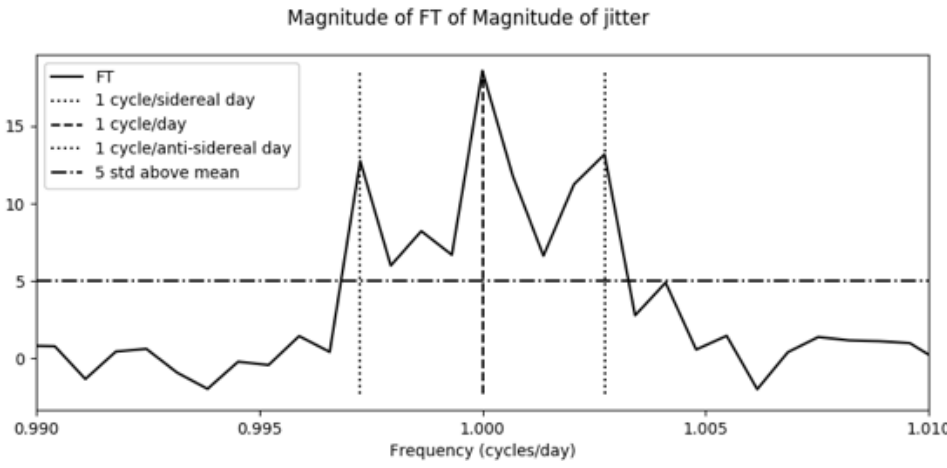

Figure 2 – Magnitude of the FT of magnitude of jitter from an actual experiment over the past two and a half years with 2 clocks. Three peaks are noticeable at one cycle per anti-sidereal day, day and sidereal day. The second graph is an enlargement around1 cycle/day.

This graph should pique our interest that something is causing a periodic variance in the jitter of our clocks. In the next section, we will propose a hypothesis that correctly predicts these spikes and allows us to in turn measure the velocity of the CMB reference frame in the process.

## 3. Hypothesis

### 3.1. Relativity

Centuries ago, Galileo argued that one cannot determine the speed of an inertial frame without looking at a reference [1]. Today, we ask if a deeper level of absolute space-time could exist beyond what has been measured and verified. What we find is an interesting hypothesis that particles move via the discrete Bernoulli process and expose a new paradigm of looking at variance as a potential source of new physics.

Specifically, we find that by conditioning the first assumption of special relativity to apply only to the mean value of the space-time coordinates of a particle or field, then the conclusions of special relativity remain the same, but we show that when we consider the variance of the Bernoulli process, there is a reference frame where the jitter is minimized.

Depicted in Figure 3, the Lorenz transformation [4] applied to the mean value of the location of a particle drifting in an inertial frame helps define the parameters of the Bernoulli process. It can be noted that we also assume the second postulate of special relativity that the speed of light is constant. Furthermore, from Dirac, we know that the eigenvalue of the speed of any particle is the speed of light [14], which sets the ratio of the step size in position to the step size in time.

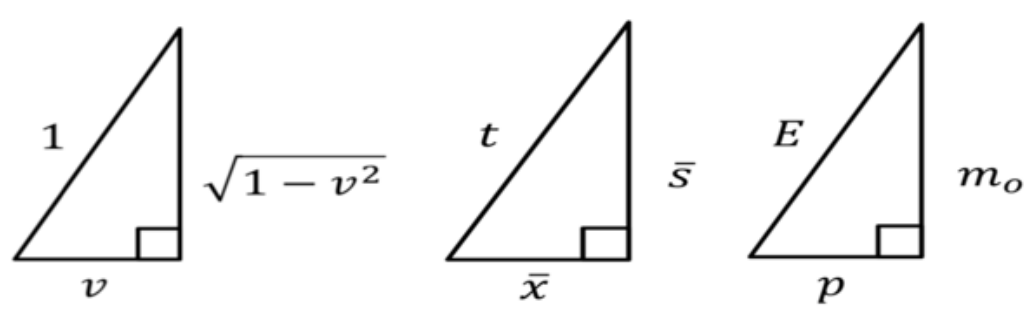

Figure 3 – The geometry of the Lorentz transformation when $c = 1$. While these triangles are similar, they are not the same. We use them to define the parameters of the Bernoulli process.

### 3.2. Bernoulli Process

While more details are available in the methods section, we will review the main points here. The Bernoulli process shows how particles step to the left or right with a certain probability. Each subsequent step is independent and identically distributed (i.i.d.) and is known as a discrete random walk. Chandrasakhar and Reif [9,10] calculate the average displacement and its variance, where $\beta$ is the probability of stepping to the right, $N$ is the number of steps and $\delta x$ is the step size. We have

$$\overline{x(N)} = (2\beta - 1)\delta x N$$

$$\left(\Delta x(N)_\beta\right)^2 = 4\beta(1-\beta)(\delta x)^2 N$$

What this is saying is that a particle does not move smoothly across space with a constant motion, but rather it jumps back and forth like zitterbewegung [15].

As stated above, the speed of these jumps is $c$, the speed of light [14]. However, it is the mean value that moves smoothly over many steps and follows $x = vt$. Now, if $\delta x = c\delta t = \hbar/2m_o c$ and $t = N\delta t$, we can see our first hint at our governing thought.

$$(2\beta - 1) = v/c$$

Notice that this equation is mathematically nice since $\beta \in [0,1]$ and $v \in [-c, c]$. However, a consequence of a particle that has biased drift, $\beta \neq 0.5$, is that the variance, which was maximized in the position dimension at $\beta = 0.5$, shifts from the position dimension to the proper time dimension, $s$. What results is a width of the possible locations of the particle after a certain time that is dependent on its velocity, see figure 4. Since no reference frame is defined, we consider this velocity to be relative to a preferred frame.

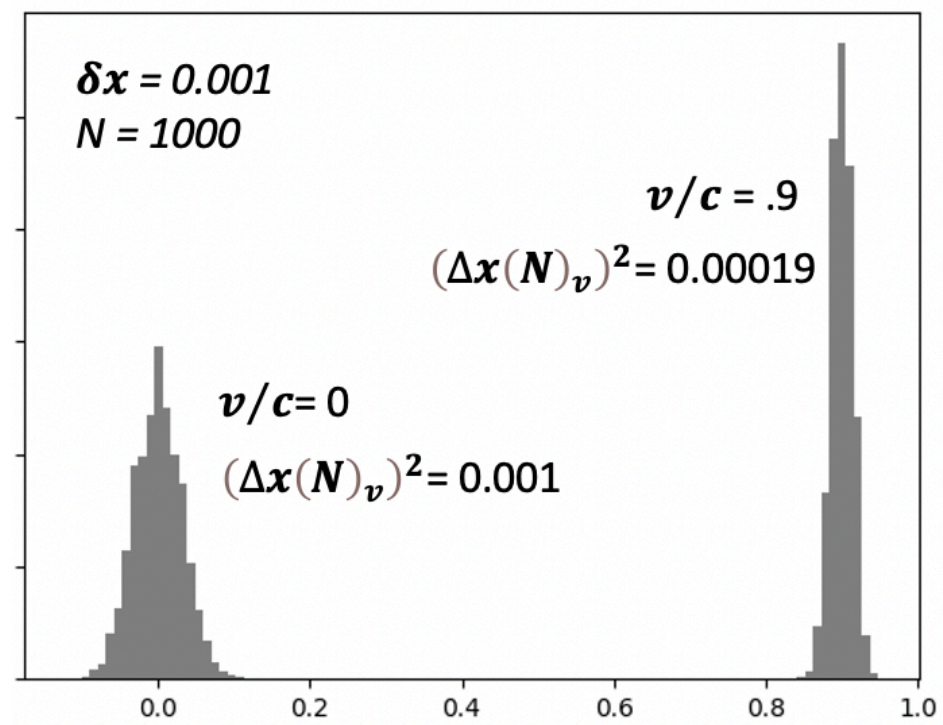

Figure 4 – Illustrative diagram of 2 different ensembles undergoing the Bernoulli process (random walk). One with an average velocity of zero and the other 0.9c. Note that the ensemble that is moving with a nonzero average velocity is thinner.

In fact, as shown in the methods section, when we use the parameters above and if we account for the impact on the variance from momentum, the variance of position follows as

$$var_r = \left|\Delta r(N)_v \hat{r} + \frac{\Delta p(N)_v t}{m}\hat{p}\right|^2 = \left(1 - \left(\frac{v}{c}\right)^2\right)\frac{\hbar t}{m}\left(1 + cos(\theta_p - \theta_r)\right)$$

and

$$var_s = \left(\frac{v}{c}\right)^2 \frac{\hbar t}{m}\left(1 + cos(\theta_p - \theta_r)\right)$$

where $cos(\theta_p - \theta_r) = \hat{r} \cdot \hat{p}$ is the inner product of the radius unit vector and the momentum unit vector.

This implication of our hypothesis breaks from the community's current understanding of relativity in that the velocity relative to the fixed starts is measurable by measuring the variance of position or proper time dimension without looking outside the laboratory. Again, it is not the 1st moment that tells the story but rather the variance.

Since relativity experiments to date have been focused on the average value of the observable or have not looked for how this varies over time, we have yet to see this effect.

### 3.3. Hypothesis applied to the laboratory

The location of a laboratory on the equator of Earth is given in Figure 5. where $f_{SR} = 1 cycle/sidereal\ day$, and $v_{z\perp}$ is the velocity of the Sun across the Universe. In the methods section, we calculate $v_{Laboratory}$ by differentiating $\overrightarrow{R_{Laboratory}}$.

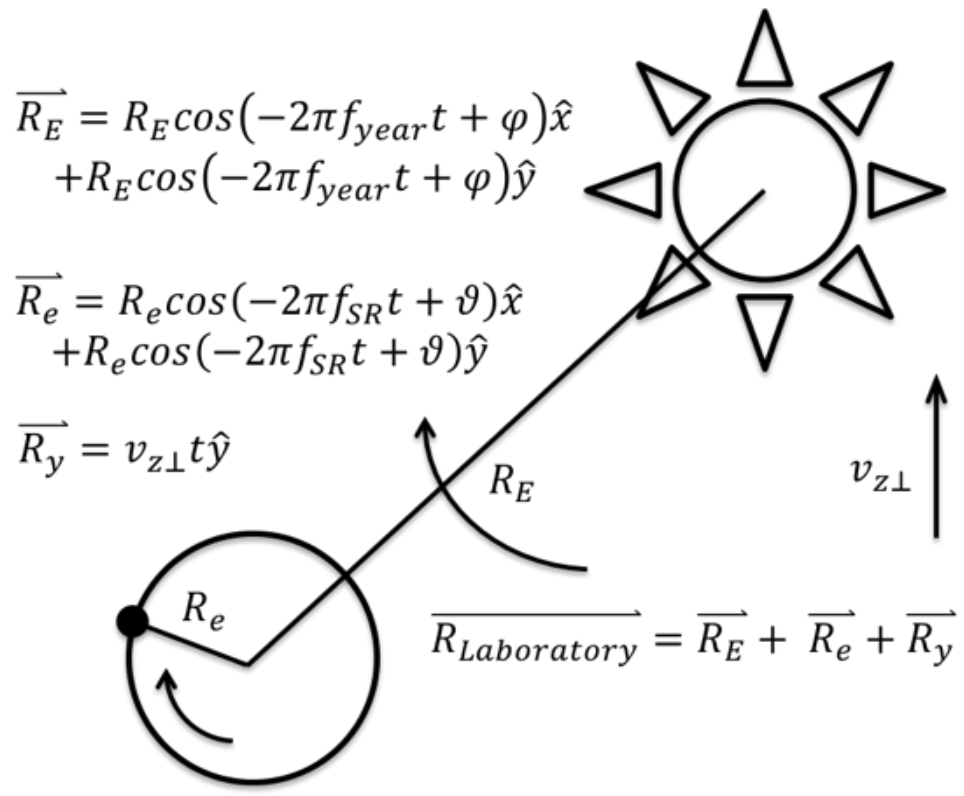

Figure 5 – A diagram of the Sun/Earth motion. There is rotation, revolution, and translational motion.

### 3.4. Implications

Putting together the implications from Sections 3.2 and 3.3, the jitter will have a term, $var_s$, where the main terms go like,

$$var_s \cong \left(\frac{|v_{z\perp}|^2}{c^2} + 2\frac{|v_{z\perp}||v_E|}{c^2} \cdot cos\left(2\pi f_{year} t - \varphi\right)\right) \cdot cos\left(2\pi f_{day} t + \varphi - \vartheta - \frac{\pi}{2}\right)\frac{\hbar t}{mc^2}$$

Here, we can see that $var_s$ has a main spike in the FT at one cycle per day and two sidebands as one cycle per anti-sidereal day and one cycle per sidereal day, just as we found in the experiments described in section 2.

## 4. Interpretation of Data

To aid in the transparency, we provide all data and code to replicate the analysis shown here [16].

If we assume that the hypothesis is correct and the spikes in the FT of the magnitude of the jitter are from the additional variance in proper time, as outlined in section 3.4, we can use the magnitudes to determine the minimum RMSE value of $v_{z\perp}$ and the phases for the minimum RMSE value of $\varphi\ and\ \vartheta$.

### 4.1. Speed

While a thorough investigation into the noise must be launched, we will consider a back of the envelop calculation. The size of the signal-to-noise ratio will have a magnitude that resembles

$$\sigma = \sqrt{\frac{var_s}{noise}} \propto \sqrt{K}$$

where $K$ is the number of samples. As outlined in the methods section, the minimum RMSE for $|v_{z\perp}|$, as indicated by the magnitude of the sidebands, leads to the conclusion that

$$|v_{z\perp}| \sim 0.001c$$

4.2. Direction

The phase calculations are straightforward. Take the phases of the three spikes and use a minimum RMSE search to determine the most likely value of $\varphi\ and\ \vartheta$.

As outlined in the methods section, both $\varphi\ and\ \vartheta$ independently point toward *Virgo.*

4.3. COSMIC Microwave Background Reference Frame

From the analysis of Planck satellite data [11], we know that in the Cosmic Microwave Background reference frame, the Sun is moving at a speed of $\sim 0.001c$ towards the boundary of $Leo$ / $Crater$. Since our experiment determined the velocity of the clocks' reference frame to within 2 hours of the CMB and has the same speed, we are encouraged to conclude that *if* our original hypothesis is correct, then the CMB reference frame is a prime candidate for the clocks' preferred frame.

What is unique about this measurement of the clocks' velocity is that we found it by looking inside the laboratory at a time standard vs. of outside at the fixed stars for a reference.

5. Information Mechanics

One might ask, what led to the fundamental question, "Is time discrete." The answer is found in an effort I am calling, *Information Mechanics* [12] and is based on the simple idea that information equals energy times time.

5.1. "Information equals Energy times Time"

Take for example, the common knowledge for over a century, that the rate of information that can be encoded into a signal is linear in the bandwidth [17,18]. Since the bandwidth is limited to the carrier frequency, $f$, and with Planck's work on black body radiation and Einstein's equation for the photoelectric effect, where $\epsilon = hf$ [20], we have the following direct proportion.

$$I \propto ft \propto \epsilon t$$

Where $I$ is the information contained in the signal, $f$ is the frequency of the particle, $\epsilon$ is energy (Planck's constant times $f$) and $t$ is the duration.

Now if we measure the energy and the time using their standard deviations and using Heisenberg's Uncertainty Principle [20] we can simply conclude that information is energy times time divided by the quantum of action.

$$I = \frac{\epsilon \mathrm{t}}{\Delta\epsilon \Delta t} = 2\epsilon \mathrm{t}/\hbar$$

5.2. Three derivations equating entropy and action

While the argument above is simple, it behooves one to go deeper. Below we give the highlights of three derivations in its true form, that entropy is equal to action. These derivations are supported with additional details in the methods section.

We have,

1) A bottom up derivation using information theoretic methods such as the gaussian channel, mutual information, and conditional entropy

2) A path integral derivation using the Schrödinger method, an assumption of the canonical distribution, the Hamiltonian of slow massive particle, and a Minkowski Transformation

3) A top down derivation using Newton's Laws of motion and the definition of an entropic force

5.2.1. Bottom Up

To start, we consider our hypothesis and the variance defined in section 3.2 as the signal in a Gaussian channel with the noise as its diffusion over the relaxation time. We will also look at the conditional entropy rate of the particle undergoing a biased random walk, as we outline in our hypothesis, that is convolved with Gaussian wavefunction. This finding gives insight into why the experiment was conducted in the first place.

5.2.1.1. Mutual information

Let's consider the mutual information between the vacuum and the particle. By using the Gaussian channel [21] the incremental channel capacity is,

$$dC = \frac{1}{2}\log\left(1 + \frac{S}{N_D}\right)$$

As we hinted to above, we take the signal as the variance of one step in the vacuum Bernoulli process (in an inertial frame when $cos(\theta_p - \theta_r) = 0$), and the noise as the diffusion of the particle over the relaxation time, $\tau = \hbar/2k_B T$,

$$S = (\Delta x)^2_{Vaccum} = \left(1 - \left(\frac{v}{c}\right)^2\right) 2(\delta x)^2 K$$

Using Einstein's kinetic relation for $D_\mu$ [22]

$$N_D = 2D_\mu \tau = 2\mu k_B T \tau = \frac{v\hbar}{F}$$

We can Taylor expand the logarithm and account for both the positive and negative energy states [14], giving a factor of 2 in the differential mutual information, $dI_M = 2dC.$ By further reducing the result as shown in the methods section, dividing by $\delta t$, and making the assumption that when the potential energy is zero, the mutual information rate is zero, then the information rate is equal to the potential energy, $V$, divided by the quantum of action.

$$\frac{\mathrm{d}I_M}{\mathrm{d}t} = \frac{2}{\hbar}V$$

From here we see the Taylor expansion is valid if $2Vdt/\hbar \ll 1$

#### 5.2.1.2. Conditional entropy

We find the conditional entropy of the particle given its interaction with the vacuum by splitting a Gaussian wavefunction. Described in detail in [12] and discussed in the methods section, the Gaussian can be separated between a positive and negative state to produce binary entropy plus an additional $log(e/2)$ found using Hirshman's approach (which is to add the differential entropy from both domains) [23].

If $\Phi(x)$ is the Gaussian wavefunction with zero mean and $\Delta x$ standard deviation, we can construct a probability distribution.

$$p_\Phi(x) = \beta|\Phi(x - \delta x)|^2 +$$

$$(1-\beta)|\Phi(x + \delta x)|^2$$

The resulting conditional entropy is the sum of the differential entropy from both domains and the binary entropy.

$$H_C(\beta) = \frac{1}{2}log(2\pi e(\Delta x)^2) + \frac{1}{2}log\left(2\pi e\left(\frac{\Delta p}{\hbar}\right)^2\right) + H_2(\beta)$$
$$= log(e/2) + H_2(\beta)$$

Where,

$$H_2(\beta) = -\beta\log(\beta) - (1-\beta)\log(1-\beta)$$

With $(2\beta - 1) = v/c$, (again from our hypothesis in section 3.2) we can Taylor expand the logarithm if $|v/c| \ll 1$. Pulling this all together as we do in the methods section and again divide by $\delta t = \hbar/2mc^2$ to get the rate, we have

$$\frac{\mathrm{d}H_C}{\mathrm{d}t} = \frac{2}{\hbar}\left(mc^2 - \frac{mv^2}{2}\right) = \frac{2}{\hbar}(mc^2 - K)$$

Because the particle has interacted with the vacuum potential to make it kinetic, we call $H_c$ the conditional entropy, $H(particle|vacuum\ potential)$. The resulting biased diffusion is a function of the state of the particle after its interaction with the vacuum.

#### 5.2.1.3. "Entropy" and "Action"

From information theory, the negative expected log probability, $H$, is equal to the conditional entropy plus the mutual information [21]. Thus, we have,

$$H = -\mathrm{E}[log(p)] = H_c + I_M$$

Furthermore, with an assumption of the canonical distribution we can show $H$ is the thermodynamic entropy over Boltzmann's constant [10],

$$H = -\mathrm{E}[log(p)] = \log(\Omega) = S/k_B$$

Thus,

$$\log(\Omega) = \left(\frac{2}{\hbar}\right)\int(mc^2 - \mathcal{L})dt$$

We now see that the self-information of the particle, or its entropy, is equal to its action over the quantum action, i.e. the time integral of the mass energy minus the Lagrangian over $\hbar/2$.

### 5.2.2. Schrödinger formulation

The central idea of this derivation is to use a Minkowski transformation to extract the phase from the wave function which otherwise, due to unitarity, would cancel positive and negative frequencies. Given the Hamiltonian,

$$\mathcal{H} = m_o c^2 + \frac{m_o v^2}{2} + V$$

And with a simple Minkowski transformation, *it→t,* [4,6,20] gives us,

$$p(dt) = |\psi(dt)|^2 = e^{-2\left(m_o c^2 - m_o v^2/2 + V\right)dt/\hbar}$$

With either use of the Asymptotic Equipartition Property [21] or an assumption of the canonical distribution [10], we have,

$$\log(\Omega) = \left(\frac{2}{\hbar}\right)\int(m_o c^2 - \mathcal{L})dt$$

### 5.2.3. Top Down

In this derivation we use a thermodynamical and mechanical approach via entropic forces, of which there is much talk today [24,25]. Using this idea, we are able to give a top down derivation of the unification of action and entropy using only: the definition of entropic forces [24], the Euler-Lagrange equations [26] and Newton's Laws of motion [2].

With more insight in the methods section, here we will start with Newton's 1st Law with a term for entropic forces and one for the back-action vacuum force, $F_{vacuum} = ma$

$$F_{entropic} + F_{conservative} + F_{vacuum} = 0$$

Using the Euler-Lagrange equations as derived from Hamilton's principle [26] and the definition of conservative [19] and entropic forces [24], we have

$$k_B T\nabla\log(\Omega) - \nabla V(r) + \nabla\left(\frac{mv^2}{2}\right) = 0$$

Integrating over the gradient and introducing the relaxation time, $\tau = \hbar/2k_B T$, we have,

$$\log(\Omega) = \left(\frac{2}{\hbar}\right)(E_o - \mathcal{L})\,\tau$$

With a constant of integration, the negative Lagrangian times time is "action" and the log of the number of states is "entropy".

## 6. Conclusion

While these tools have provided evidence of a reference frame where the jitter is minimized, we emphasize that all the conclusions of special relativity remain the same and are consistent with the above if we modify the first postulate of special relativity to be conditioned on the "mean" value of the space-time coordinates, but allow for the variance to bring new insight.

### 6.1. Additional experiment underway

The source of the signal we found in the jitter experiment is fundamentally diffusion. We suggest that this diffusion is also present in Johnson noise on a resistor. We are currently in the process of measuring the magnitude of the noise on three resistors spanning the Cartesian coordinates.

Since Johnson noise is proportional to the variance of momentum (which is proportional to temperature), we just need to adjust this variance due to the edge effects of the binomial distribution. From section 7.2.1, with where, $N = mc^2/k_BT_o$, we have,

$$var_{p_x} = \left|\Delta p_x \widehat{p_x} + m\frac{\Delta x}{t}\hat{x}\right|^2 = \left(1 - \left(\frac{v_x}{c}\right)^2\right) 2k_BT_o m(1 + \hat{x}\cdot\widehat{p_x})$$

Since $\hat{x}\cdot\widehat{p_x} = 0$, this means the effective temperature is,

$$T_{eff} = \left(1 - \left(\frac{v_x}{c}\right)^2\right) T_o$$

And thus, with $\mathcal{B}$ the bandwidth, the voltage noise is,

$$(\Delta V)^2 = 4k_BT_oR\mathcal{B}\left(1 - \left(\frac{v_x}{c}\right)^2\right)$$

With $v_x = |v_{z\perp}|cos(2\pi \cdot f_{SR}t + \zeta)$, we have

$$(\Delta V)^2 = 4k_BT_oR\mathcal{B}\left(1 - \left(\frac{|v_{z\perp}|}{c}\right)^2 \cdot \frac{(1 + cos(2\pi \cdot 2f_{SR}t + 2\zeta))}{2}\right)$$

With the signal to noise ratio looking like

$$SNR = -\left(\frac{|v_{z\perp}|^2}{c^2} + 2\frac{|v_{z\perp}||v_E|}{c^2}\cdot cos(2\pi f_{year}t - \varphi)\right) \cdot \frac{(1 + cos(2\pi \cdot 2f_{SR}t + 2\zeta))}{2}$$

Which has frequencies we can easily search for at 2 cycles per sidereal day with sidebands at $\pm$ one cycle per year.

### 6.2. Unification

It's possible we have just scratched the surface of what is waiting to be explored. Both "action" and "entropy" are each a foundation of significant disciplines in physics. Specifically, for action we can derive of the Laws of Motion, both Newton's and Einstein's, follow the evolution of the quantum state by maximizing or minimizing the path action integrals, and define the magnitude of spin. And for entropy we can determine the arrow of time, measure the disorder of a system, and explore information theory and thermodynamics [19].

A unification of these two cornerstones could yield unexpected and previously unseen advances as did the other proven unifications like Newton's unification of gravity and astronomy in the 17th century, Maxwell's unification of electricity magnetism in the 19th century, Einstein's unification of space-time / mass-energy in the 20th century and Quantum Mechanics unification of fields and particles also in the 20th century.

For example, what if we consider the quantum of action as a quantum of information what could that tell us about the nature of particles or the vacuum. Could particles be simple Turing machines that generate random numbers and move based on how those numbers compare to a code stored in the vacuum [27]? Could the desire to unify the forces benefit from an insight such as the area of a spin network being [12,13],

$$A = \frac{4G\hbar}{c^3}\sum_i (s_i + 1/2)$$

As a last note, the implications even go beyond physics in the debate about absolutism vs relativism. In philosophy and theology, the concept of absolutism has been waning. Could these insights help swing pendulum back in the other direction?

Acknowledgements

JLH would like to thank Robert Jahn, formerly of Princeton University, and Stephen Hawking, formerly of Cambridge University, who in 1997 and 1998 in separate private conversations contributed to and verified the original thought (information equals energy times time) respectively.

JLH is also grateful to: his wife and two boys for the reminder about what is most precious; Ron & Freddy of GuideTech for providing the A2D jitter analyzers at cost; Gage for customizing the digitizers, SpectraTime for discounting the atomic clocks; Bemco for customizing and discounting the environmental chamber; Rajesh Kumar Prabhakaran for setting up the cloud based and on-site high performance compute clusters; Jiyan Pan for being a sounding board on the analysis; a council of colleagues and friends who have helped shaped the story arc, and all the other friends and family who shared in this journey.

## 7. Method

In this extended section we review more details necessary to replicate these findings and support the original derivation of the governing thesis. We will share these details backwards.

### 7.1. Theory

To augment the three derivations in the main section, we will give more details to fill in the blanks.

#### 7.1.1. Bottoms Up derivation

##### 7.1.1.1. Mutual information

We start where we left off above,

$$\mathrm{d}I_M = 2\mathrm{d}C = \frac{2\left(1 - \left(\frac{v}{c}\right)^2\right) Fc^2(\delta t)^2}{v\hbar}$$

We now make the interpretation that $\delta t$, is the time differential $\mathrm{d}t$ and $v = dx/dt$. Making these two changes and divide by $\mathrm{d}t$,

$$\frac{\mathrm{d}I_M}{\mathrm{d}t} = \frac{2Fc^2}{\frac{\hbar \mathrm{v}}{\mathrm{d}t}} - \frac{2Fv\mathrm{d}t}{\hbar}$$

Replacing $F = mv/\mathrm{d}t$ and $v\mathrm{d}t = \mathrm{d}x$; we next consider conservative forces so we can replace the incremental work, $F\mathrm{d}x$, with the negative of the potential energy, $V$, minus the reference potential, $V_0$.

$$\frac{\mathrm{d}I_M}{\mathrm{d}t} = \frac{2}{\hbar}(mc^2 + V - V_0)$$

To solve for $V_0$, we postulate that the particle and the vacuum are independent when the potential is zero, $V = 0$. Thus,

$$\frac{2}{\hbar}(mc^2 - V_0) = 0$$

When the two are independent, the mutual information rate is zero [21],

And thus

$$\frac{\mathrm{d}I_M}{\mathrm{d}t} = \frac{2}{\hbar}V$$

#### 7.1.1.2. Conditional entropy

Here we show how a Gaussian wavefunction layered on top of the Bernoulli Process generates $log(e/2)$ additional to the binary entropy [12]. We take the Gaussian wave function,

$$\Phi(x) = \frac{1}{\sqrt[\frac{1}{4}]{2\pi(\Delta x)^2}} e^{\frac{-(x)^2}{4(\Delta x)^2}}$$

If we convolve $\Phi(x)$ with a Dirac delta function impulse pair [28] (biased by $\beta$) the resulting probability distribution is,

$$p_\Phi(x) = \beta|\Phi(x - \delta x)|^2 + (1-\beta)|\Phi(x + \delta x)|^2$$

If $\delta x > \Delta x$, then the differential entropy of the resulting probability distribution is

$$\frac{1}{2}log(2\pi e(\Delta x)^2) - \beta\log(\beta) - (1-\beta)\log(1-\beta)$$

We can similarly compute the Fourier Transform. Due to the phase modulation properties, the resulting probability distribution in the wavenumber space is the same without splitting $\Phi(x)$ into two. Thus, the differential entropy is

$$\frac{1}{2}log\left(2\pi e\left(\frac{\Delta p}{h}\right)^2\right)$$

As instructed by Hirshmann [23], when we add the differential entropy of these two Fourier Transform pairs, we get

$$log\left(e\frac{\Delta x \Delta p}{\hbar}\right) - \beta\log(\beta) - (1-\beta)\log(1-\beta) =$$

$$log(e/2) + H_2(\beta)$$

With $(2\beta - 1) = v/c$, (again from our hypothesis in section 3.2) we can Taylor expand the logarithm if $|v/c| \ll 1$

$$H_2(\beta) = \log(2) - \frac{\left(\frac{v}{c}\right)^2}{2}$$

Thus, the total conditional entropy of the particle is

$$H_C(\beta) = H_2(\beta) + \log\left(\frac{e}{2}\right) = 1 - \frac{\left(\frac{v}{c}\right)^2}{2}$$

in natural units per $\delta t = \hbar/2mc^2$. Bringing the conditional entropy rate to

$$\frac{\mathrm{d}H_C}{\mathrm{d}t} = \frac{2}{\hbar}\left(mc^2 - \frac{mv^2}{2}\right) = \frac{2}{\hbar}(mc^2 - K)$$

### 7.1.2. Schrödinger formulation

While it is possible to derive the relationship that $I = -log(p)$ from the Asymptotic Equipartition Property (AEP) [12], we can also get to the same result by assuming the canonical distribution [10].

$$\frac{S}{k_B} \equiv \log(\Omega) = -E[log(p)]$$

And thus, from the concavity of the log we have [21],

$$-E[log(p)] = -log(p)$$

Now we use the Schrödinger formulation of quantum mechanics to solve the wave equation, where $\mathcal{H}$ is the Hamiltonian and is equal to the imaginary time derivative operator, with the energy eigenvalue being the rest mass energy plus the kinetic energy plus the potential energy [20],

$$\mathcal{H}\psi = \frac{i\hbar d}{dt}\psi = \left(m_o c^2 + \frac{m_o v^2}{2} + V\right)\psi$$

$$\psi = e^{-i\left(m_o c^2 + m_o v^2/2 + V\right)t/\hbar}$$

From the principles of quantum mechanics [19,20] we know,

$$p(dt) = |\psi(dt)|^2$$

Which results in $p(dt) = 1$, leading to zero information. However, if we are willing to consider breaking the unitary requirement, a simple Minkowski transformation where $it \to t$ [4,6,20] gives us,

$$p(dt) = |\psi(dt)|^2 = e^{-2\left(m_o c^2 - m_o v^2/2 + V\right)dt/\hbar}$$

Notice that the kinetic energy is negated upon the Minkowski transformation. With

$$p = \prod p(dt)$$

We have,

$$\log(\Omega) = \left(\frac{2}{\hbar}\right)\int (m_o c^2 - \mathcal{L})dt$$

### 7.1.3. Top Down

In this methods section we will give insight into how we negate $ma$. Let's start with Newton's 2nd Law of Motion and include terms for conservative forces and entropic forces,

$$F_{entropic} + F_{conservative} = ma_{particle}$$

If we proceed from here, the result without does not negate the kinetic energy. We can solve this issue by again using the Minkowski transformation in this derivation. However, we will choose to provide a little more insight into the vacuum's role in this by negating the acceleration through Newton's 3rd Law.

By considering the vacuum-particle system and consider the acceleration of the vacuum, $a$, from Newton's 3rd Law, we have

$$ma_{particle} = F_{particle} = -F_{vacuum} = -ma$$

We can thus re-write the above equation as a statement of Newton's 1st Law with the acceleration from the perspective of the vacuum,

$$F_{entropic} + F_{conservative} + F_{vacuum} = 0$$

Using the Euler-Lagrange equations as derived from Hamilton's principle we have [27]

$$F_{vacuum} = ma = \frac{d}{dt}(mv) = \nabla\left(\frac{mv^2}{2}\right)$$

We also know from mechanics that [19]

$$F_{conservative} = -\nabla V(r)$$

And with the definition of entropic forces we have [24]

$$F_{entropic} \equiv T\nabla S = k_B T \nabla \log(\Omega)$$

Putting this together we have

$$k_B T \nabla \log(\Omega) - \nabla V(r) + \nabla\left(\frac{mv^2}{2}\right) = 0$$

Integrating over the gradient and shifting the Lagrangian to the other side of the equation, we have

$$k_B T \cdot \log(\Omega) = V - \frac{mv^2}{2} + E_o = E_o - \mathcal{L}$$

Where $E_o$ is the constant of the integration. Now by introducing the relaxation time, $\tau = \hbar / 2k_B T$, we have

$$\log(\Omega) = \left(\frac{2}{\hbar}\right)(E_o - \mathcal{L})\,\tau$$

## 7.2. Hypothesis

### 7.2.1. Bernoulli Process

Above we gave a brief explanation of how the parameters of the Bernoulli Process fit the relationships in figure 4. Let us fill in.

If we set the probability that time increments by a time step, $\delta t = \hbar / 2m_o c^2$, equal to one. Thus $\bar{t} = N\delta t$, where $N$ is the number of steps. Since t has no variance (as its probability parameter is one) we can set $\bar{t} = t$.

Next for $x$, when $\delta x = c\delta t$, we have,

$$\bar{x} = (2\beta - 1)N\delta x = (2\beta - 1)ct$$

In this case, since the probability parameter is neither zero nor one, x will have a variance and thus we cannot rewrite $\bar{x}$ *as* $x$. Here we see the velocity eigenvalue is equal to the speed of light, as Dirac found [14,15], but since we know that the mean displacement of the particle is equal to the average velocity times time, we have

$$(2\beta - 1) = v/c$$

Likewise, we have the variance of position follows as,

$$\left(\Delta x(N)_\beta\right)^2 = 4\beta(1-\beta)(\delta x)^2 N$$

Plugging in the equation above between $v$ and $\beta$

$$(\Delta x(N)_v)^2 = \left(1 - \left(\frac{v_x}{c}\right)^2\right)(\delta x)^2 N$$

We also need to consider the momentum space

$$\overline{p_x} = (2\beta_x - 1)mc$$

Which leads to

$$(\Delta p_x(N)_v)^2 = \frac{4\beta_x(1-\beta_x)(mc)^2}{N} = \frac{\left(1 - \left(\frac{v_x}{c}\right)^2\right)(mc)^2}{N}$$

A calculation of variance across all three spatial dimensions is,

$$\begin{aligned} var_r &= \left|\Delta r(N)_v \hat{r} + \frac{\Delta p(N)_v t}{m}\hat{p}\right|^2 \\ &= (\Delta r(N)_v)^2 + \left(\frac{\Delta p(N)_v t}{m}\right)^2 \\ &\quad + 2\,cos\left(\theta_p - \theta_r\right)\Delta r(N)_v \frac{\Delta p(N)_v t}{m} \end{aligned}$$

$$var_r = \left(1 - \left(\frac{v}{c}\right)^2\right)\frac{\hbar t}{m}\left(1 + cos\left(\theta_p - \theta_r\right)\right)$$

And for proper time, we have,

$$var_s = \left(\frac{v}{c}\right)^2 \frac{\hbar t}{m}\left(1 + cos\left(\theta_p - \theta_r\right)\right)$$

Where $cos\left(\theta_p - \theta_r\right) = \hat{r} \cdot \hat{p}$

### 7.2.2. Laboratories velocity

In figure 5 we outline the location of the laboratory on Earth's equator. We calculate the velocity by differentiating, which results in,

$$\begin{aligned} \left|v_{Laboratory}\right|^2 = {} & |v_{z\perp}|^2 + |v_E|^2 + |v_e|^2 \\ & + 2|v_{z\perp}||v_E| cos\left(-2\pi f_{year} t + \varphi\right) \\ & + 2|v_{z\perp}||v_e| cos(-2\pi f_{SR} t + \vartheta) \\ & + 2|v_E||v_e| cos\left(-2\pi f_{day} t + \vartheta - \varphi\right) \end{aligned}$$

We can also find $cos\left(\theta_p - \theta_r\right)$ for our laboratory since the dominating momentum (which determines $\hat{p}$) is $2\pi \cdot mass \cdot R_E f_{year}$ in the direction of Earths revolution, while the dominating centrifugal acceleration (which determines $\hat{r}$) points out from the center of the Earth. Thus,

$$cos\left(\theta_p - \theta_r\right) = cos\left(2\pi f_{day} t + \varphi - \vartheta - \frac{\pi}{2}\right)$$

### 7.3. Experiment

Here we will review:

- the data flow
- the equipment used in the jitter experiment
- the processing used to remove the phase noise
- the results of the analysis
- the direction of the velocity
- the method used for the LIGO strain

#### 7.3.1. Data Flow

The origin of the signal is two rubidium clocks that are each put through a square wave circuit to sharpen the edges. Next an analog to digital converter with two channels measures the zero crossings. Each channel timestamps are differentiated to get the time periods and the difference between channel 0 and channel 1 is computed. This time series, sampled at 1.25MHz, is stored as a binary data file every 15 seconds and transferred to 1 of 2 compute machines.

In the next processing, the script loads the binary data, takes another derivative to remove drift and the absolute value is computed. Note that it is essential to take the absolute value, just like when you rectify an amplitude-modulated signal before low pass filtering to get the encoded envelope.

This time series is a proxy for the variance of the jitter. Next for a select set of 161 frequencies centered at 1cycle/day with a $\Delta f = 1\text{cycle}/4\text{years}$, the Fourier Transform of the time series is computed for each 15 second data file and the 161 complex coefficients are stored in a database. At the same time the Fourier Transform of the unit function is calculated and the coefficients are likewise stored. Then every hour the coefficients from each 15 second data file are aggregated and stored for long term analysis. Since we are tallying the results in real time (due to the experiment producing 1TB of data daily) the output of the Fourier Transform is similar to a spectrogram with a complex value stored for each hour for each frequency.

#### 7.3.2. Equipment

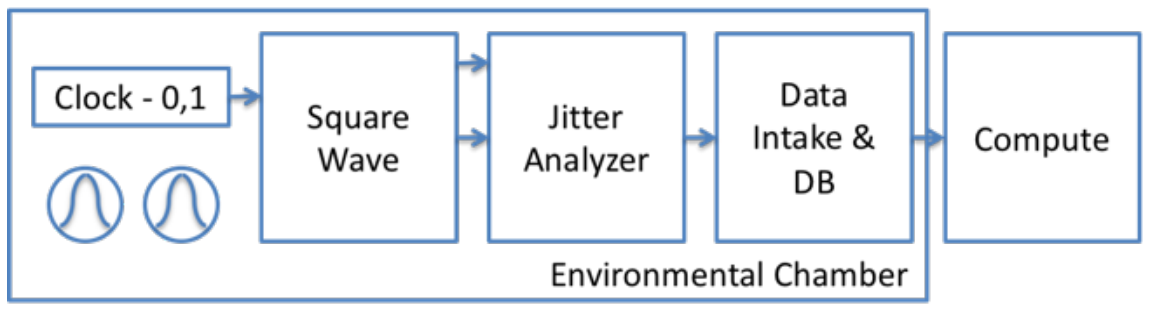

Figure 7 – the physical flow of the apparatus to measure the variance of jitter on two rubidium clocks.

The specific equipment includes (See Figure 7): 2 LPFRS Rb atomic clocks from Spectratime, 2 LTC6957-1 square wave converters, a gt668 jitter analyzer from GuideTech, an i5 Dell Inspireon to host the analyzer and DB, and 2 i7 Dell Inspireons to process the Fourier Transform. The clocks and jitter analyzer are contained in a Bemco environmental chamber. The temperature and humidity of the A2D converter is measured and stored every minute and available with the rest of the data and code to replicate this analysis on Google Drive [16].

For each frequency, the only processing is to subtract a scaled and smoothed FT of the unit function to remove the phase noise (see Figures 7 and 8) and then sum the spectrogram coefficients over time. This noise cancelation processing helps compensate for the fact that some data are missing or low quality and that each of the select frequencies are not exactly full cycles. This process follows the guidance found in Bracewell [28].

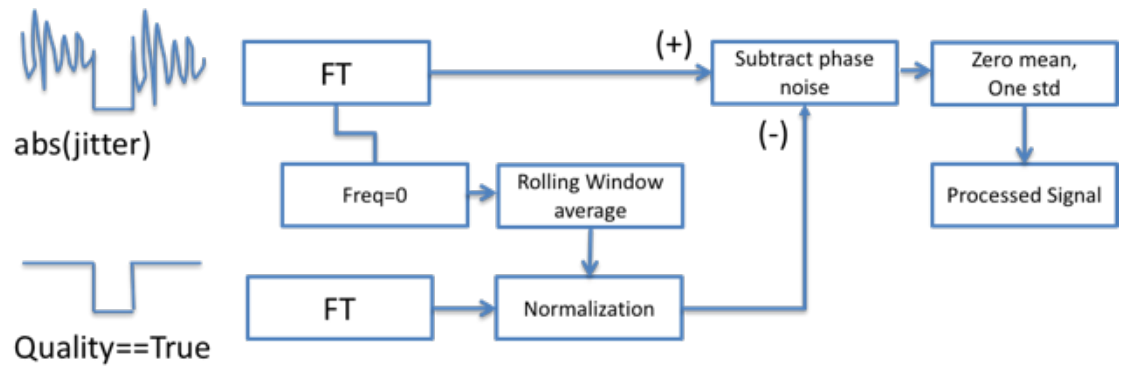

Figure 7 – Flow diagram of postprocessing analysis. We can see that when data are missing or the quality is poor, we compensate for what would be a large impact from the phase noise. This ensures that we see the smaller signal that only shows itself through many coherent hours. Last, we normalize to zero mean and one standard deviation.

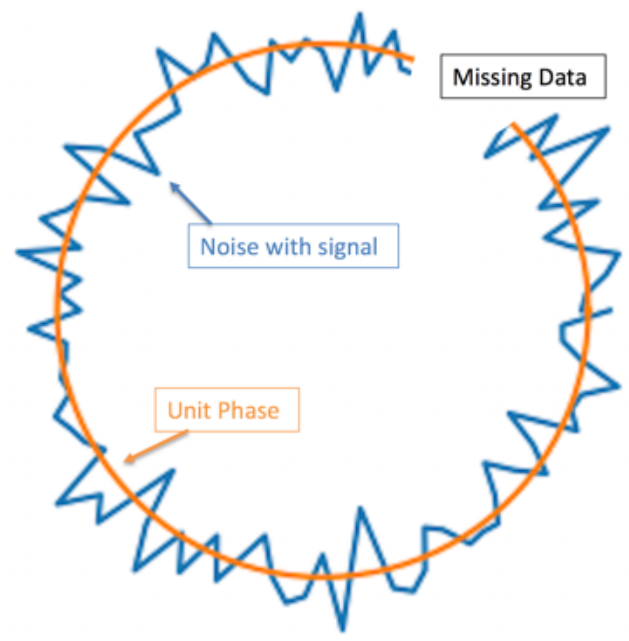

Figure 8 – Illustration of phase noise cancellation – during the times of bad or missing data, we preserve the slight aggregate phase impact of the signal over the cycle. When data are not missing, we subtract a constant full cycle, which is zero.

In the end, we sum over the duration of the experiment and have the aggregate coefficients in the frequency domain. Finally, we normalize the frequency coefficients and graph the magnitude, which results in Figure 2 above.

#### 7.3.3. Results

The peaks of the graph in Figure 2 are at:

$$12.6\ \sigma\ @ -3.07\ radians\ @\ f_{ASR}$$

$$18.4\ \sigma\ @\ 2.37\ radians\ @ f_{day}$$

$$13.0\ \sigma\ @\ 1.96\ radians\ @\ f_{SR}$$

#### 7.3.4. Speed

Considering the setup of the experiment, we will have the size of the spikes in the FT of the magnitude of the jitter that will look like below. Note that this is not the final assessment of the noise but rather a starting point.

$$\boldsymbol{\sigma_{1/day}} \approx \sqrt{\frac{\frac{1}{2} \cdot \frac{1}{2} \cdot \frac{|v_{z\perp}|^2}{c^2} \left(\frac{\hbar \Delta t}{mc^2}\right)}{\eta^2} 2K}$$

$$\boldsymbol{\sigma_{1/day \pm 1/year}} \approx \sqrt{\frac{\frac{1}{2} \cdot \left(\frac{1}{2}\right)^2 \cdot \frac{2|v_{z\perp}||v_E|}{c^2} \left(\frac{\hbar \Delta t}{mc^2}\right)}{\eta^2} 2K}$$

Here, $K$ is the number of samples, $\Delta t$ is the period of an individual sample (in this case, $1/1.25MHz$), $m$ is the mass of the electron and $|v_E|$ is the speed of the Earth around the Sun. σ will be similar to the square root of one half of the signal since we measured the absolute value, not the variance, times one half per split of the power between positive and negative frequencies, divided by the noise power (in this case, $\eta = 6e - 12sec$) and multiplied by $2K$ since we have two clocks, each with $K = 9.7x10\text{^}13$.

We calculate the minimum RMSE for $|v_{z\perp}|$ when we consider the measured value of $\sigma_{1/day \pm 1/year}$ and the equation above.

$$|v_{z\perp}| \sim 0.001c$$

However, the measured $\sigma_{1/day}$ is lower than expected if we use $|v_{z\perp}| = 0.001c$ in the equation above. This is likely due to a variation in temperature also at $1cycle/day$ but out of phase, as shown in section 6.2.7. Thus, this negative correlation likely reduces the signal at $1cycle/day$, and accounting for this effect could lead to the conclusion being the same about $|v_{z\perp}|$.

### 7.3.5. Direction vector

For the direction of the velocity, we have 2 degrees of freedom $\varphi$ *and* $\vartheta$ and three datapoints (the phases of the three frequencies). By minimizing the RMSE as shown in the analysis scripts [16], the best fit values for $\varphi$ *and* $\vartheta$ are:

$$\varphi = 0.62 \; radians \; and \; \vartheta = 2.83 \; radians$$

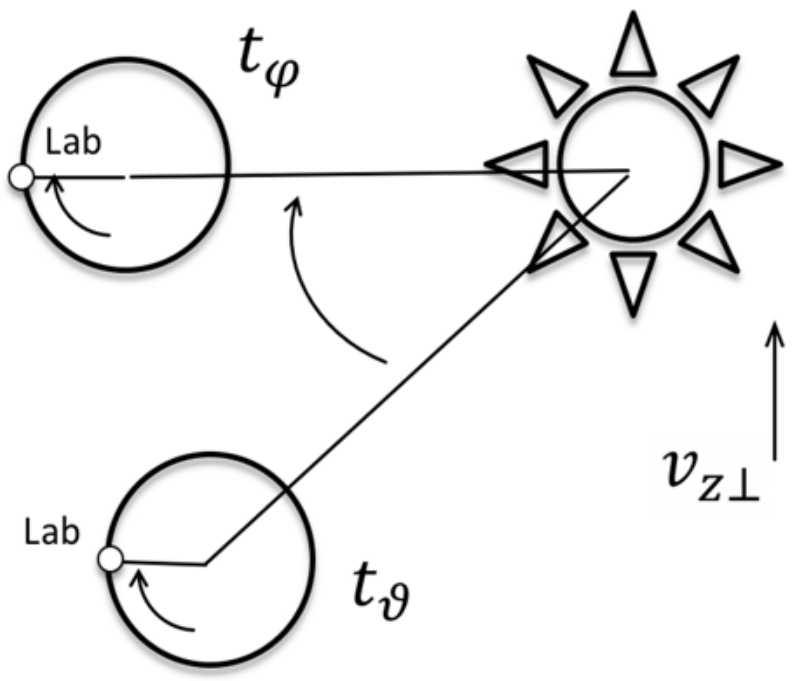

Figure 9 – Two phase (direction) parameters of the RSME best fit data suggest the Laboratory is moving close to, if not in the same direction as the CMB reference frame.

To find the direction of the velocity of the Sun in the preferred frame, we start with the equation $\varphi = 2\pi f_{year} t_\varphi$ to find $t_\varphi$~3,156,000. On that date, our velocity is pointed toward the *Virgo* on the Eastern Horizon at solar midnight local time in St George, UT (the location of the laboratory) [29,30]. See figure 9. Similarly, we can confirm this direction by using $\vartheta$ and its associated unix time $t_\vartheta = 38{,}893$, when it so happens *Virgo* is again on the Easter Horizon in St George, UT on Jan 1 1970 [29]. *Virgo* is immediately next to the *Leo-Crater* border.

### 7.3.6. Ruling Out Other Sources

One must question these results and attempt to consider other sources of the signal. For example, temperature has a natural frequency of $f_{day}$ and might be the source of the signal. However, there is evidence that this influence can be ruled out in the jitter experiment as the cause.

Specifically, the entire experiment (the clock and the A2D converter) was shielded in an environmental chamber that drove the overall correlation in FT space between the temperature and magnitude of jitter to 63% but with a phase of $\sim\pi$. Thus, when the temperature peaked, the magnitude of the jitter was in a trough. Furthermore, from a visual inspection the temperature does not have sidebands at $\pm 1cycle/year$ like the jitter does. Additionally, the correlation with humidity is 3%.

Furthermore, if the temperature impacts the jitter by increasing the noise when the temperature is larger, and since the temperature and jitter Fourier Coefficients are out of phase at $f_{day}$, we could consider the actual $\sigma_{1/day}$ to be larger. With this effect, the main spike could also be consistent with $|v_{z\perp}| \sim 0.001c$.

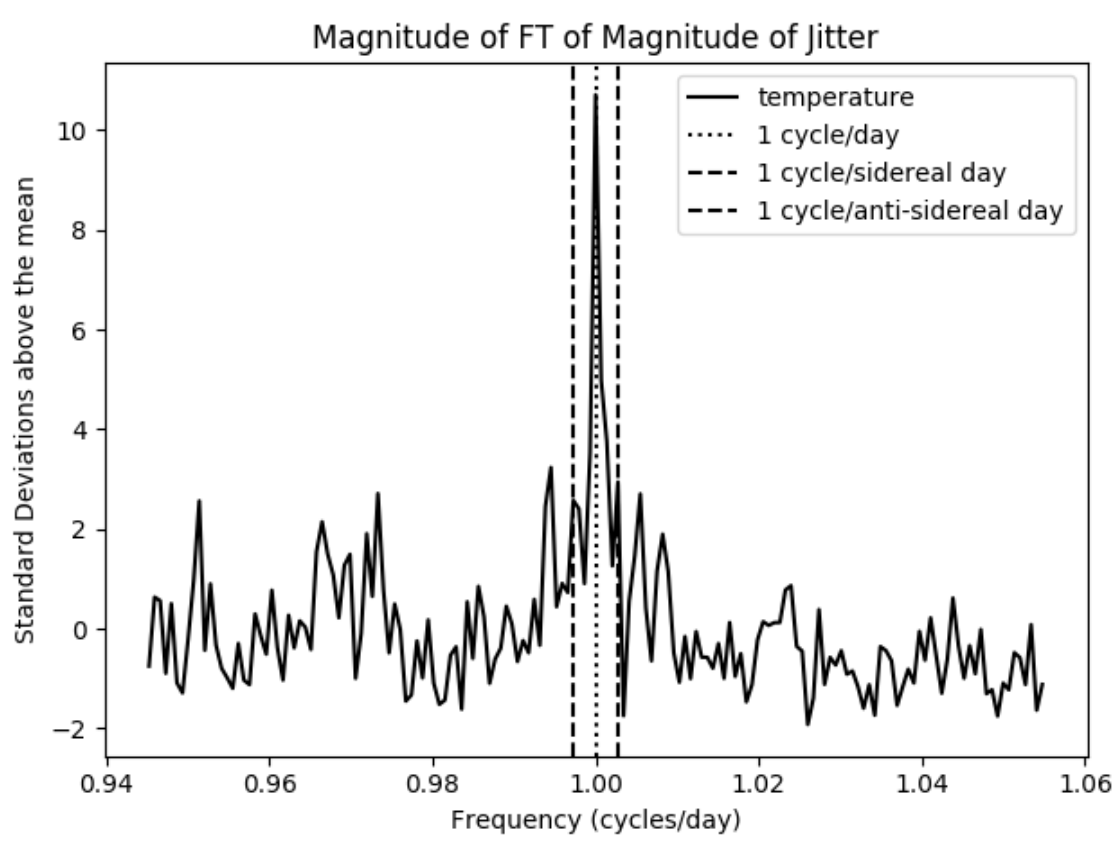

Figure 10 – The Fourier Transform of the temperature of the clocks and A2D converter. Despite the environmental chamber, a spike at $f_{day}$ is prevalent; yet this spike is completely out of phase with the jitter at $f_{day}$, and the coefficients at $f_{anti-sidereal\ day}$ and $f_{sidereal\ day}$ are below the noise floor.

We see that the temperature is not a driver of the two sidebands, as shown in Figure 10. One might consider cosmic rays to be the cause; however, we do not see signal at 367.25 cycles per year and out proposed experiment to measure diffusion in space will confirm this.

#### 7.3.7. Other Experimental Environments

There are other classes of experimental environments where one can see the artifacts of the Bernoulli process. The first class is at the individual step and the difference between the ensemble distribution and the first moment. An ensemble of particles will move discretely based on probability, it's only over many steps that the Law of Large Numbers kicks in and its average velocity follows special relativity. This class of experiments is likely to be found in extremely low-temperature environments such as a 2D electron gas.

A second class of experiments is looking at the variance over many steps, as in the jitter experiment outlined in this article, or at other times, keeping standards across the world. The position dimension would also show a signal where the wavefunction will spread as a function of the particle's orientation.

A third class of experiments is an interferometer. While Michelson and Morley (and subsequent experiments) did not find variation in the speed of light relative to orientation, a similar experiment, but one that measures variance instead of the mean value, might show a signal. Since position variance is dependent on the 3D orientation, there will be more spikes in the signal and will be shifted by 2 cycles per sidereal day.

LIGO is an excellent example of such an experiment; however, with the low duty cycle and low bandwidth of the measurements, there is just not enough data to show a distinctive SNR. However, an attempt was made to find this signal, and for anyone interested, the code to download and analyze the data is available on Google Drive [16]. Nevertheless, an interferometer that is optimized around Brownian noise should show this signal.

JLH is grateful for the LIGO collaboration, as this material is based upon work supported by NSF's LIGO Laboratory, which is a major facility fully funded by the National Science Foundation.